\documentclass[12pt]{article}

\usepackage{graphicx}

\begin{document}   
\title{Critical Exponents of the KPZ Equation via 
Multi-Surface Coding Numerical Simulations}

\author{
Enzo Marinari$^{(1)}$, Andrea Pagnani$^{(2)}$ and 
Giorgio Parisi$^{(1)}$\\[0.5em]
\small $^{(1)}$: Dipartimento di Fisica, INFM and INFN, 
      Universit\`a di Roma {\em La Sapienza}  \\
\small P. A. Moro 2, 00185 Roma, Italy\\
\small $^{(2)}$: Dipartimento di Fisica and INFM, Universit\`a di Roma 
         {\em Tor Vergata}\\
\small Viale  della Ricerca Scientifica 1, 00133 Roma , Italy\\[0.3em]
\small e-mail: {\tt Enzo.Marinari@roma1.infn.it, 
Andrea.Pagnani@roma2.infn.it}\\
\small {\tt Giorgio.Parisi@roma1.infn.it}
}

\date{May 2000}

\maketitle
 
\begin{abstract}
\noindent 

We study the KPZ equation (in $D$ = $2$, $3$ and $4$ spatial
dimensions) by using a RSOS discretization of the surface.  We measure
the critical exponents very precisely, and we show that the rational
guess is not appropriate, and that $4D$ is not the upper critical
dimension. We are also able to determine very precisely the exponent
of the sub-leading scaling corrections, that turns out to be close to
$1$ in all cases.  We introduce and use a {\em multi-surface coding}
technique, that allow a gain of order $30$ over usual numerical
simulations.

\end{abstract}

\section{Introduction}

The KPZ equation \cite{KPZ}, in its apparent simplicity, involves many
issues that need clarification.  The continuum equation (that
describes the local growth of an interface profile) is

\begin{equation}
 \label{kpz_equation}
  \frac{\partial h}{\partial t} =
  \nu \vec{\nabla}^2 h
  + \frac{\lambda}{2} \left(\vec{\nabla} h\right)^2
  + \eta(\vec{r},t)\ .
\end{equation}
The real behavior of this equation is not well understood. The absence
of a complete mean field theory does not help, and the fact that we
have do understand the effects of a strong coupling fixed point makes
very difficult the development of a perturbative renormalization
approach.

A practical approach for numerical studied of the problem is to
consider lattice Restricted Solid On Solid (RSOS) surfaces (see for
example \cite{RMP} for a review): one considers a discretized height
field on a $D$ dimensional lattice, and imposes the constrain that the
absolute value of the distance among two neighboring surface elements
can only take the values zero and one.

Old extensive numerical simulations \cite{TANGETAL,ALANISETAL} do not
clarify the situation completely.  After a number of interesting Field
Theoretical results \cite{FT}, the introduction of new Renormalization
Group based techniques is a potentially promising direction of
development \cite{CAMAPI}.

The main theoretical points which still deserve proper explanation are
two: it is still not clear whether equation (\ref{kpz_equation}) has a
finite upper critical dimension ($D_>$) or not, and the exact
quantification of the related critical exponents. Regarding arguments
in favor of $D_>=4$ we address the reader to \cite{LASKIN}, while
arguments supporting $D_>=\infty$ can be found in \cite{CAMAPI,YU}.

In this light we introduce here a new numerical technique and present
some precise numerical simulations that allow us to estimate critical
exponents. Thanks to the new, precise technique we succeed to clarify
some questions: we show for example that a conjecture of
\cite{RATIONAL} is not founded, and we give quantitative estimates
about the behavior of sub-leading corrections. 


\section{The Multi-Surface Coding}

The precise results on which this paper is based are due to a new
technique we introduce to simulate RSOS surfaces, where adjacent
elements of the surface can only be at a distance of $1$, $0$ or $-1$
lattice spacings. The technique is a generalization of the so called
{\em multi-spin coding} technique (well discussed by Rieger
\cite{RIEGER} in the context of the study of disordered systems),
where, by using the fact that the $\pm 1$ spins can be represented by
Boolean logical variables, one stores $64$ copies of the system in one
single computer word (we will assume in the following we are using a
$64$ bit computer). The main idea is that we can simulate, basically
at the cost of one single usual simulation, $64$ copies of the system,
by rewriting the basic operations (like summing spins for computing
the effective force) as boolean operations, and by exploiting the fact
that when, for example, the computer is calculating a logical bit
$AND$ it is indeed doing that $64$ times at once. The gain of such an
approach over an usual single spin simulation is of order $30$: one
gains a factor $64$ for the number of systems that updates at once,
and looses a factor of order $2$ in computational complexity (adding
the spin in the Boolean form takes some more time \cite{RIEGER}).

We generalize here the Ising case to the case of a {\em field of
differences}, that can have three values, since a given element of the
surface can be at the same height of a given neighbor, or one step
behind it, or one step ahead of it. The method is new as applied to
such a system, where one does not have to compute the value of an
energy, but uses the boolean operations to determine if the element
can be moved without violating the geometrical constraint.

Let us thinks for simplicity about $2D$, where we have a $2D$ support
where a height field (the surface height) $h_i=h_{x,y}$ is
defined. One surface element $h_i$ has $2D=4$ first neighbors: in the
RSOS model the difference $ \Delta_{i,\mu} \equiv h_i -
h_{i\pm\hat{\mu}}$, where $\mu=x,y$, can only take the three values
$-1$, $0$ and $+1$.  We store the $DL^2$ values of the
$\Delta_{i,\mu}$ (and, as we will discuss better, each
$\Delta_{i,\mu}$ needs two bit to be stored.  This is a redundant way
to store the information, since a factor two in memory could be easily
saved, but it is very convenient from the point of view of computer
time: as usual one trades memory for time, and avoids a cumbersome
reconstruction by storing more information).

If a given element is behind its neighbor the difference is $-1$, if
it has the same height than the neighbor the difference is $0$, while
if it is ahead of the neighbor the difference is $+1$. Since we have
three allowed values we can represent each of these difference with
two bits (that we will call respectively $H$, high bit, and $L$, low
bit): we can for example code with $00$ the situation where the given
element is behind its neighbor, with $01$ the situation where they
have the same height, and with $10$ the situation where it is ahead
(the value $11$ is forbidden).

If the element is ahead of even a single one of the $2D$ neighbors the
move is forbidden (it would violate the RSOS constrain). It is clear
this is easily implemented in our coding: one just needs to do a
logical $OR$ of the $2D$ $H$ bits related to the site $i$, and if it
is $1$ at least one of the neighbors is behind and the move is not
allowed. Let us see better why. We start comparing our element $i$ to
the first of its $2D=4$ neighbors (let us say the one in the positive
direction $+1$): if $i$ is ahead the neighbor the relevant $H$ bit
(that we have stored in $\Delta_{i,+1}$) is $1$, and we already know
we cannot move. If $i$ is on the contrary at the same height of the
neighbor or behind it we have that, as far as this neighbor is
concerned, the element $i$ can advance of one unit. Now we look at the
second neighbor, where the same reasoning holds: if we look at the
logical $OR$ of the two relevant high bit we will find that we cannot
move if this quantity is one. Looking at all the $2D$ neighbor we see
that the move is forbidden if the logical $OR$ of the $2D$ $H$ bit is
$1$: clearly this operation, as all the other present in the core of
the code, updates with a single computer cycle the $64$ copies of the
system. After doing that, if the element $i$ cannot be moved there is
nothing that has be done: if it gets moved we have to update the
$\Delta_{i,\mu}$ related to the site $i$ and to the neighbors, to
describe the new situation.

The codes simulating systems in a different number of spatial
dimensions (in our case from $D=2$ to $D=4$) are simple generalization
one of the other (when adding a dimension one has to add checks on the
new neighbors and their update).

As usual in these kind of simulations the random noise is implemented
with a random choice of the site to be updated.  We also implement a
fifty percent probability of really updating a surface element that
according to the RSOS constrain could be updated.  This is very
important, since starting from random independent surfaces is not
enough: because of our parallel scheme the sites of our $64$ copies
have to be updated in the same order, and such an updating algorithm
with updating probability one is attractive, and the $64$
configurations asymptotically at large times become equal
\cite{LIGGET}. The probability of not accepting an allowing change,
that depends on each of the $64$ configurations, solve this problem.

We are aware of a parallel algorithm to update surfaces \cite{FORTAN}:
it is very different in spirit from our algorithm (since it
parallelizes on different sites of the lattice). We have not compared
in detail the performances of the two algorithms, but we believe that
on one side the algorithm of \cite{FORTAN} is more general, and not
limited to RSOS models, but on the other side our algorithms is more
regular (there are no exceptional loops), and is probably better
performing for the model we study.

\section{The Numerical Simulation}

We have based this work on numerical simulations of lattices of
volume $V\equiv L^D$.  The spatial dimensionality $D$ is the spatial
support, where a $1$ dimensional surface take values.  We study the
$D=2$, $D=3$ and $D=4$ cases.  At a given time $t$ of the dynamical
evolution the position of the surface can be expressed by the values
$h_i(t)$ (that we reconstruct by the differences we store in our code,
see the former section).  $i$ represents, in lexicografic order, a
$D$-ple labeling the spatial sites.

We consider a dynamics which generates a Restricted Solid on Solid
(RSOS) growth: distances of first neighboring elements of the surface
cannot be larger than one. At each trial step we move elements of the
surface that are not constrained not to do so because of the RSOS
restriction with probability $\frac12$: we have to use a probability
different from one in order to keep independent the different surfaces
we simulate in the same computer word.

We consider a large number of different lattice sizes. In $D=2$ we
take $L$ going from $5$ to $641$, in $D=3$ we consider $L$ going from
$5$ to $103$ while in $D=4$ $L$ goes from $5$ to $28$.  All our data
have been averaged over $64$ different dynamical runs.

\begin{table}
\centering
\begin{tabular}{|c|c|c|c|c|c|}
\hline
$L(2D)$ & $T(2D)$ &$L(3D)$ & $T(3D)$ &$L(4D)$ & $T(4D)$ \\
\hline
5     & 7.50002e+06  &   11 & 7.5e+06      &   7 & 786450\\   
\hline
41    & 7.50002e+06  &   13 & 7.5e+06      &  11 & 1.2e+06\\
\hline
79    & 6.00002e+06  &   17 & 7.5e+06      &  13 & 1.2e+06\\
\hline
157   & 6.891e+06    &   19 & 7.5e+06      &  15 & 1.2e+06\\
\hline
317   & 1.6475e+06   &   23 & 7.5e+06      &  17 & 898050\\
\hline
641   & 233070       &   25 & 3.74905e+06   &  20 & 510150\\
\hline
      &              &   33 & 1.4542e+06    &  22 & 523650\\
\hline
      &              &   37 & 1.04265e+06  &  25 & 786450\\
\hline
      &              &   61 & 484350       &  28 & 456150\\
\hline
      &              &   83 & 696600       &     & \\
\hline
      &              &  103 & 409050       &     & \\
\hline
\end{tabular}
\caption[1]{
Number of full lattice sweeps for each run on different lattice sizes
and number of dimensions.
\protect\label{TAB-ONE}}
\end{table}

Because of the way we use to determine critical exponents, by trying
to measure precisely the asymptotic time behavior, we use very long
runs, and we always try to check that we have reached the asymptotic
plateau in a clear way (see the discussion and the figures in the next
section).  Let $t$ be the time labeling sweeps of our simulation (we
define a sweep as the trial update of $V$ random sites).  In any
of our simulations we run $T$ updating sweeps: we give in table
\ref{TAB-ONE} the number of full lattice sweeps for each run on
different lattice sizes and number of dimensions.  We measure the
observable every $1000$ lattice sweeps: when analyzing the large time
asymptotic behavior of the system we discard the first half sweeps (we
call $T_0$ the time of our first measurement): this is a very
conservative attitude, but we prefer to be safe on not having any
systematic bias by paying a price of making maybe the statistical
error ten percent larger of the best we could do.

We define the time dependent observables

\begin{equation}
  \overline{h(t)} \equiv \frac1{V} 
  \sum_{i=1}^V \left[ h_i(t) \right]\ ,
\label{E-HBART}
\end{equation}
and

\begin{equation}
  w_k(t) \equiv \frac1{V} 
  \sum_{i=1}^V \left[ \left( h_i(t) - 
  \overline{h(t)} \right)^k \right]\ ,
\label{E-WKT}
\end{equation}
that we compute for $k=2$, $3$ and $4$ and for different $D$ and $L$
values (when needed we will label $w$ with the upper script $(L)$, to
make clear to which lattice size we are referring). We define the
large time asymptotic limit of $w_k(t)$ as

\begin{equation}
  w_k^{(L)} \equiv \frac{1}{T-T_0+1}\sum_{t=T_0}^T
  w_k(t)\ ,
  \label{E-WKL}
\end{equation}
for simulation on a lattice of linear size $L$. We always check (and
this is one of the crucial points of this note, that we will discuss
in better detail in the next section) that $T_0$ and $T$ are large
enough to make our result unbiased in the precision of our statistical
error.

\section{Analysis}

We will discuss here the analysis of our numerical data.  We want to
determine the critical exponents of the asymptotic behavior of the
$w_k$ we have defined in the former section.  For example we have that
the asymptotic infinite time value of $w_2$ has a leading scaling
behavior

\begin{equation}
  w_2^{(L)} = L^{2 \chi}\ ,
  \label{E-CHI}
\end{equation}
while at intermediate times (large enough for being in the scaling
region but  small enough not to feel the finite size of the lattice)

\begin{equation}
  w_2(t) = t^{\frac{2\chi}{z}}\ .
  \label{E-ZETA}
\end{equation}
The first
behavior is obtained by taking large times on different lattice sizes,
and by studying the time asymptotic value as a function of $L$. The
second behavior is studied by simulating large lattices, and analyzing
the behavior of the systems for times larger than one, but very
smaller than the thermalization time (at the given value of the
lattice size). An exact (Galilean) invariance of the KPZ equation
implies that $z+\chi=2$. Deciding if it is better to measure
accurately $z$ or $\chi$ is a practical matter.

As opposed to the choice, for example, of reference \cite{ALANISETAL},
here we have mainly based our analysis on fitting the behavior of the
large time asymptotic value $w_k^{(L)}$, and we have only used fits
to the intermediate time behavior to substantiate our results. We
believe in fact that it is very difficult to determine a precise
quantitative estimate of the exponent of the time scaling.
The problem with the time dependent behavior of $w_k(t)$ is that, in
order to get an unbiased value, one needs a double cutoff, both at
small and at large time. At small times the behavior of $w_k(t)$ is not
a pure power, and one has to discard small lattice, and/or to use
corrections to scaling, in order to remove this effect. At large times
one starts to feel the finiteness of the lattice system, and a new
crossover (toward the asymptotic, constant time behavior) intervenes.
In  other words a careful analysis of the time exponent needs a double
sliding window, moving both at small and at large time. We also find
that the crossover effects at large time are very important: even on
large lattices one can soon see systematic effects on the exponent
estimate due to the finiteness of the lattice.

On the contrary the time asymptotic behavior only needs one cutoff,
that excludes small times, where the asymptotic value has not yet been
reached. This is easy, and we do it by using a logarithmic division of
our data. We are in this way able to check with very high precision
that we are computing an unbiased (effective, $L$-dependent) exponent.
Again, we will also show results obtained from a direct fit of $z$, to
show they are consistent with the $\chi$ values we determine.

Our main analysis is done by fitting at the same time the three sets
of data \cite{BFMM}:

\begin{eqnarray}
\nonumber
  w_2 &\simeq& A_2 L^{      2  \chi} \left( 1 + B_2
  L^{-\omega}\right)\ ,\\ 
\protect\label{E-3FIT}
  w_3 &\simeq& - A_3 L^{      3  \chi} \left( 1 + B_3
  L^{-\omega}\right)\ ,\\ 
\nonumber
  w_4 &\simeq& A_4 L^{      4 \chi} \left( 1 + B_4
  L^{-\omega}\right)\ .
\end{eqnarray}
We always compare this fit to the best fit of $w_2$ alone (typically
by only including the large volumes and by ignoring scaling
corrections) and check that things are coherent.  We have also check
independently that, for example, $w_3$ really scales as
$w_2^{\frac32}$: this is clearly true for our data.  Without the use
of all our set of data that we have defined in (\ref{E-3FIT}) we would
not have been able to determine $\omega$ with a reasonable statistical
precision.

\begin{figure}
\centering
\includegraphics[width=0.7\textwidth,angle=0]{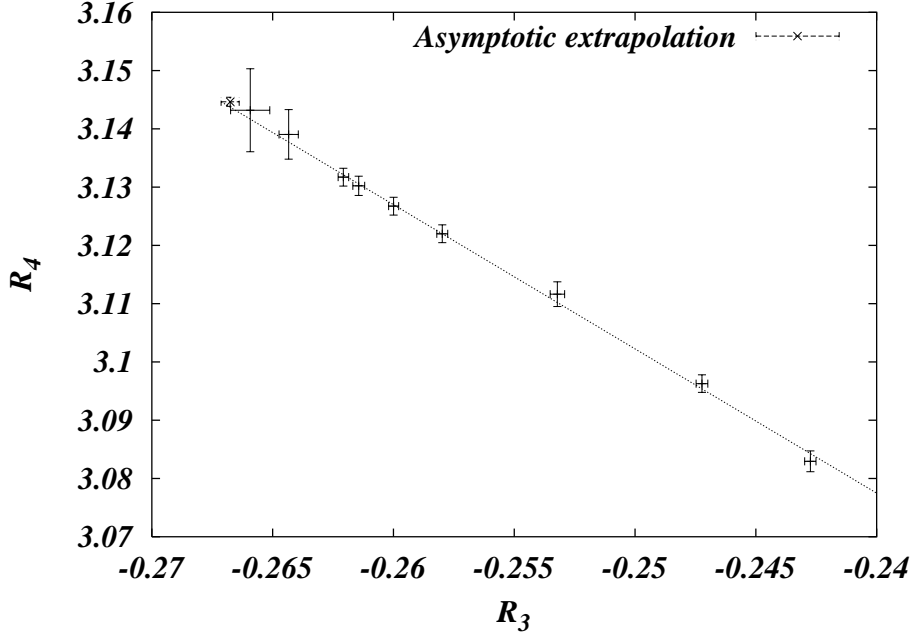}
\caption[a]{
$R_4$ versus $R_3$, $D=2$.
\protect\label{FIG-R3R4}}
\end{figure}

With this definition the exponent $\chi$ is the same of $\chi$ of
\cite{ALANISETAL} and of $\alpha$ of \cite{CAMAPI} (our definition of
dimensionality of the system excludes the time dimension, and is
always only the dimension of the space). 

The error analysis is done by using a jack-knife approach
\cite{KERKON}: we divide our statistical sample in $10$ parts all
including all of data but one tenth (each part excludes a different
tenth of the data), we fit ten time the behavior of, for example,
(\ref{E-3FIT}), and compute the error on $A_k$, $B_k$, $\chi$ and $\omega$
by the fluctuations of the ten results (multiplying the error times a
factor $10$, due to the fact that the individual parts we have formed
are correlated \cite{KERKON}). In short, we present a reliable
estimate of the statistical errors over the quantities we determine. 

The first point we have to discuss is about the exponents of finite
size corrections that appear in equation (\ref{E-3FIT}). Do the
corrections to even and odd momenta really scale with the same
exponent? This has been questioned in \cite{CHINIJ}, and we provide
here accurate evidence that $\omega_2 = \omega_3 = \omega_4 \equiv
\omega$ in our model. 

Let us start from arguing that we are not in a situation in which the
scaling exponent of the odd moments, $\omega_{odd}$ is {\em
smaller} than the scaling exponent of the even moments,
$\omega_{{even}}$ (this is the opposite scenario of the one
proposed in \cite{CHINIJ}).

Following \cite{CHINIJ} we define

\begin{equation}
R_3 \equiv \frac{w_3}{w_2^{\frac32}}\ \ ,\ \ \ 
R_4 \equiv \frac{w_4}{w_2^{2}}\ .
\end{equation}
The exponent $\chi$ disappears from these ratios, and, on general
ground, if $\omega_{{odd}}<\omega_{{even}}$, 
one has that asymptotically for large $L$, ignoring
sub-leading corrections,

\begin{equation}
R_3 \simeq c_3 + d_3\ t^{\omega_{ {odd}}}\ \ ,\ \ \ 
R_4 \simeq c_4 + d_4\ t^{\omega_{{even}}}\ ,
\end{equation}
i.e.

\begin{equation}
  R_4 \simeq c + d \left(R_3+c\right)^{\omega_{even}/\omega_{odd}}\ ,
\end{equation}
and $R_3$ is a linear function of $R_4$ if and only if
$\omega_{odd}\ge \omega_{even}$ (in which case the two ratios
asymptotically scale with the same exponent).  We plot in figure
\ref{FIG-R3R4} $R_4$ versus $R_3$, and notice that the linearity is
impressive. In these data there is no sign of a discrepancy among
scaling exponents of odd and even momenta, and they surely exclude
that $\omega_{odd}< \omega_{even}$. The last point on the left in the
figure is our asymptotic extrapolation, with attached the (small)
statistical error (the best estimate of \cite{CHINIJ} is $R_3=-0.27\pm
0.01$ and $R_4=3.15\pm0.02$, well compatible with our data but with a
very larger error).

\begin{figure}
\centering
\includegraphics[width=0.7\textwidth,angle=0]{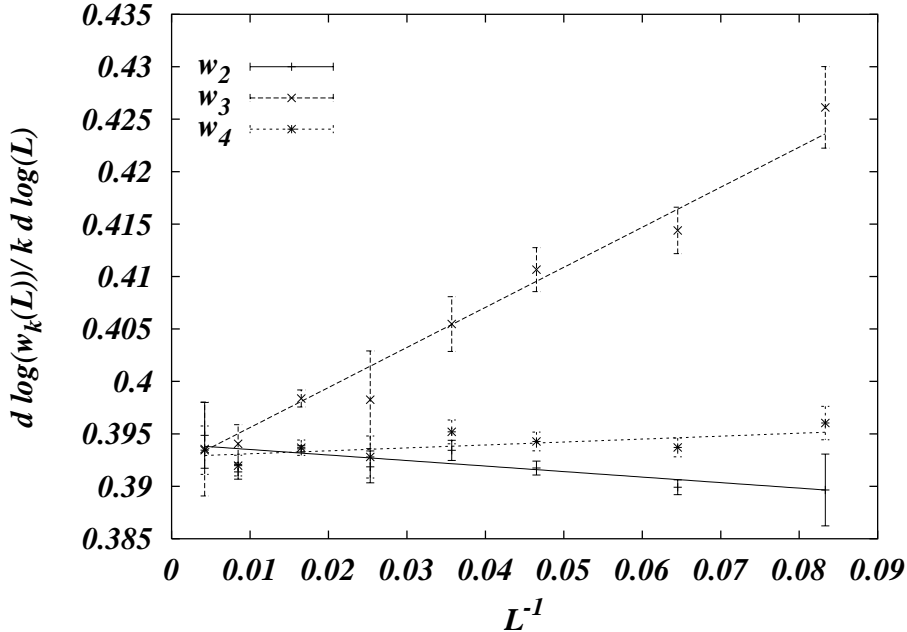}
\caption[a]{ $L\frac{d\log(w_{k}(L))}{k\,dL}$ versus $\frac1{L}$ in
$2D$, for $k=2,3,4$.  
\protect\label{FIG-LOGDER}}
\end{figure}

We use now figure \ref{FIG-LOGDER} to also exclude the case
$\omega_{odd}> \omega_{even}$, establishing in this way that for our
model $\omega_{odd}= \omega_{even}$: in figure \ref{FIG-LOGDER} we
plot the effective scaling exponent obtained by using separately the
data for $w_2(L)$, $w_3(L)$ and $w_4(L)$.  The three quantities do all
depend linearly, with very good approximation, on $\frac1{L}$, showing
that we are having the same exponent of the scaling corrections.
Again, the impressive linearity of the data implies that we are
measuring precisely a single exponent of finite size corrections, and
also that we are not mislead by finite size corrections.

Let us now discuss the determination of the exponent $\chi$.

\begin{figure}
\centering
\includegraphics[width=0.7\textwidth,angle=0]{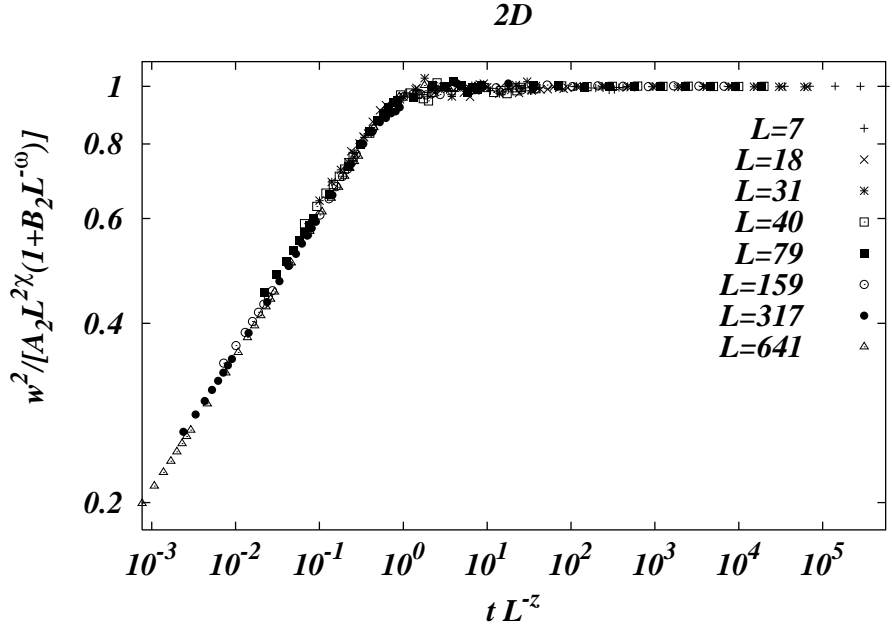}
\caption[a]{
Scaling plot of the rescaled $w_2$ versus the rescaled time. $2D$.
\protect\label{FIG-SCA2D}}
\end{figure}

\begin{figure}
\centering
\includegraphics[width=0.7\textwidth,angle=0]{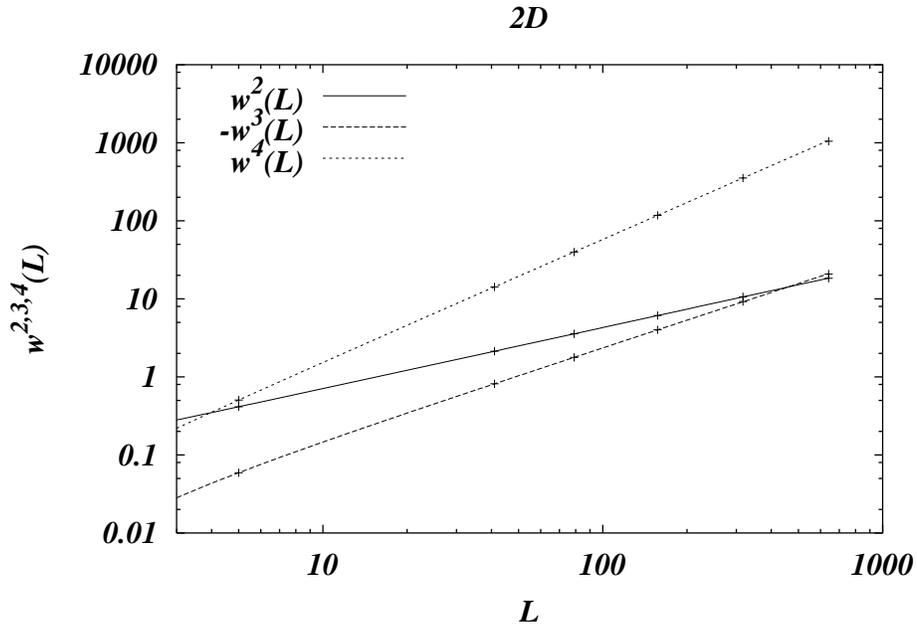}
\caption[a]{
Numerical data for $w_2$, $w_3$ and $w_4$ versus $L$ and best fit.
$2D$.
\protect\label{FIG-TRE2D}}
\end{figure}

Let us start by discussing the $D=2$ case.  A simple analysis of
$w_2^{(L)}$ without corrections to scaling shows that a fit to
lattices of linear size from $L=19$ give a good value of the chi
squared and $\chi=0.393$. A systematic analysis to the form
(\ref{E-3FIT}) by including lattice with $L\ge 11$ gives our final best
value of

\begin{equation}
  \chi_{D=2} = 0.393 \pm 0.003\ ,\ \  \omega_{D=2} = 1.1\pm 0.3\ .
\end{equation}
We plot the rescaled $w_2$ versus the rescaled time in figure
\ref{FIG-SCA2D}: it is clear that the asymptotic plateau is exposed
with good accuracy, and that the scaling is very good.  Also the best
fit to the form (\ref{E-3FIT}) is very good: we plot the numerical
data for $w_2$, $w_3$ and $w_4$ versus $L$ and the best fit in
\ref{FIG-TRE2D}.

We can compare to the rational guess of \cite{RATIONAL} that would
give here $\chi_R=0.4$. Indeed Kim and Kosterlitz in \cite{RATIONAL}
conjecture that $\chi(d)=\frac2{d+3}$ (that seemed to fit reasonably
the numerical results available at the time).  Here L\"assig
\cite{LASSIG}, by using an operator product expansion, also find
$\chi(d=2)=0.4$.  Our result is at three standard deviations from the
rational guess, that is a safe distance. Still, since we are dealing
with a very complex situation, with many corrections that can possibly
affect the result (sub-sub-leading corrections, short time, small
volume,...), we perform a further check to determine if $\chi_R=0.4$
is a plausible result. We fix $\chi=0.4$, and fit our data with now
seven and not eight free parameters ($A_2$, $A_3$, $A_4$, $B_2$,
$B_3$, $B_4$, and $\omega$). Now we get a very small value of
$\omega\simeq .28$, and a chi squared that increases of a factor $10$
from our previous best fit (where it was of order one per degree of
freedom). 

In order to show that the fact that we have been able to exclude that
the exponent takes the value $0.4$ is not due to the hypothesis that
the exponent of the sub-leading correction is the same for all momenta
we can look again at figure \ref{FIG-LOGDER}.
Already from these data it is clear that the value $0.4$ is
excluded. The more sophisticated analysis which we have presented
before is crucial in obtaining a controlled extrapolation to
$L=\infty$, keeping the statistical errors under control.  We believe
that this is a very strong evidence against the validity of the
rational guess. We will see that for higher $D$ values we get an even
clearer discrepancy.

\begin{figure}
\centering
\includegraphics[width=0.7\textwidth,angle=0]{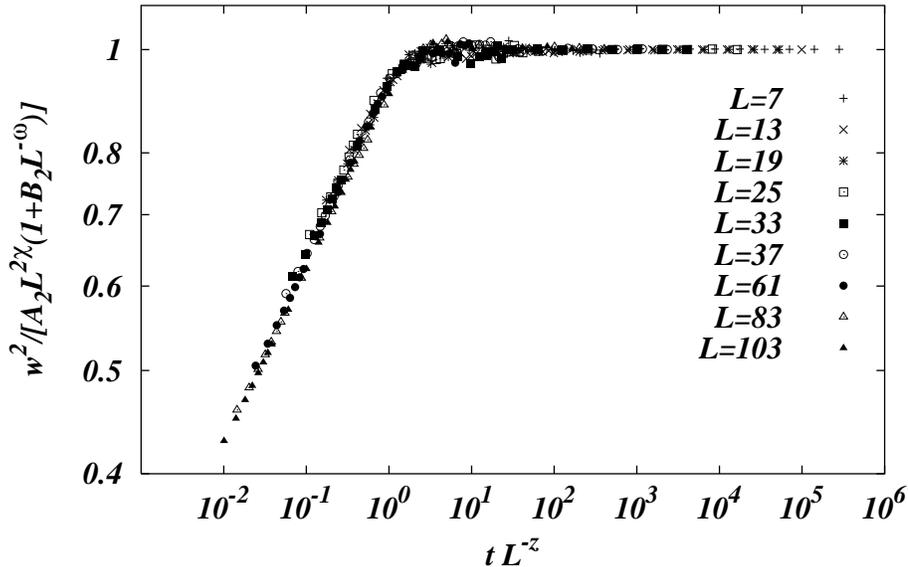}
\caption[a]{
As in figure \ref{FIG-SCA2D}, but $3D$.
\protect\label{FIG-SCA3D}}
\end{figure}

\begin{figure}
\centering
\includegraphics[width=0.7\textwidth,angle=0]{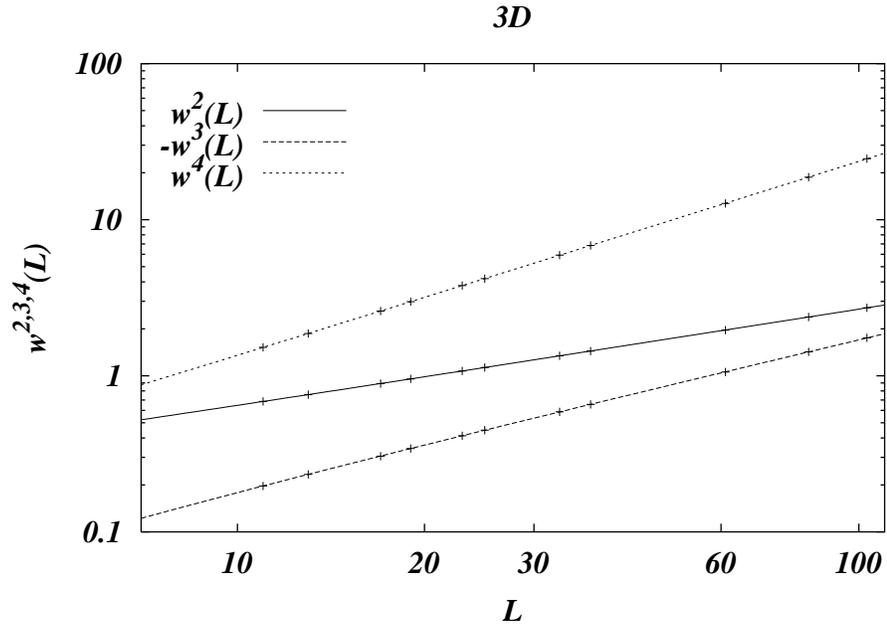}
\caption[a]{
As in figure \ref{FIG-TRE2D}, but $3D$.
\protect\label{FIG-TRE3D}}
\end{figure}

In $3D$ the same analysis of the three cumulant allows to establish
that

\begin{equation}
  \chi_{D=3} = 0.3135 \pm 0.0015\ ,\ \  \omega_{D=3} = 0.98\pm 0.08\ .
\end{equation}
We are including here sizes from $L=11$ up to $L=103$.  We plot the
rescaled $w_2$ versus the rescaled time in figure \ref{FIG-SCA3D}
and the numerical data for $w_2$, $w_3$ and $w_4$ versus $L$ and the
best fit in \ref{FIG-TRE3D}.  Here the rational guess would give
$\chi_R=\frac13\simeq 0.333$, and we are sitting at more than ten
standard deviations from it (L\"assig gives here $\frac27$ that is far
from our estimate). The same check we have done in $D=2$ leads to a
strong evidence: when fixing $\chi=\frac13$ we find again a very small
value $\omega\simeq 0.2$, and the chi squared increases of a factor
large than $20$. Here the evidence against the validity of a rational
guess is even stronger than in $D=2$.

\begin{figure}
\centering
\includegraphics[width=0.7\textwidth,angle=0]{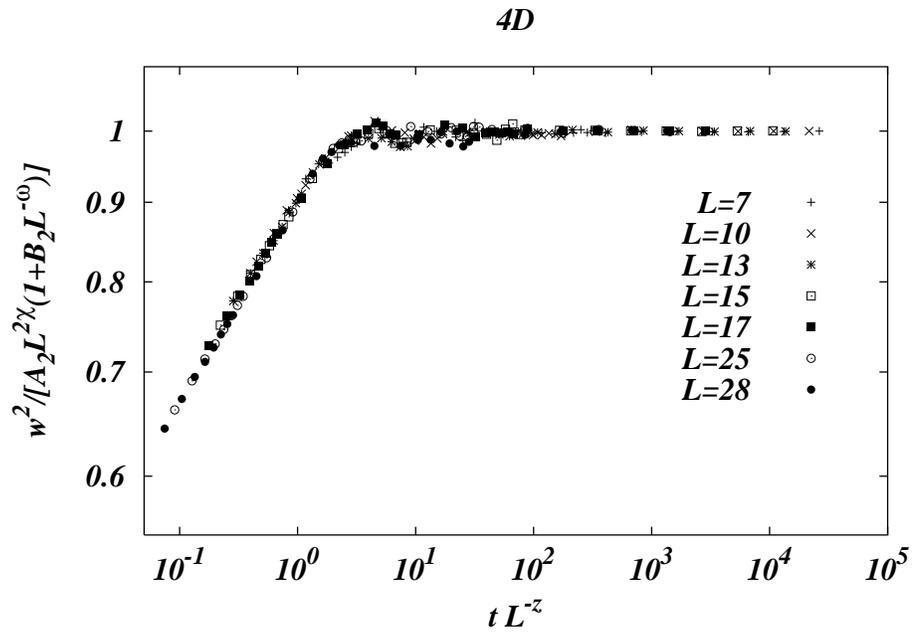}
\caption[a]{
As in figure \ref{FIG-SCA2D}, but $4D$.
\protect\label{FIG-SCA4D}}
\end{figure}

\begin{figure}
\centering
\includegraphics[width=0.7\textwidth,angle=0]{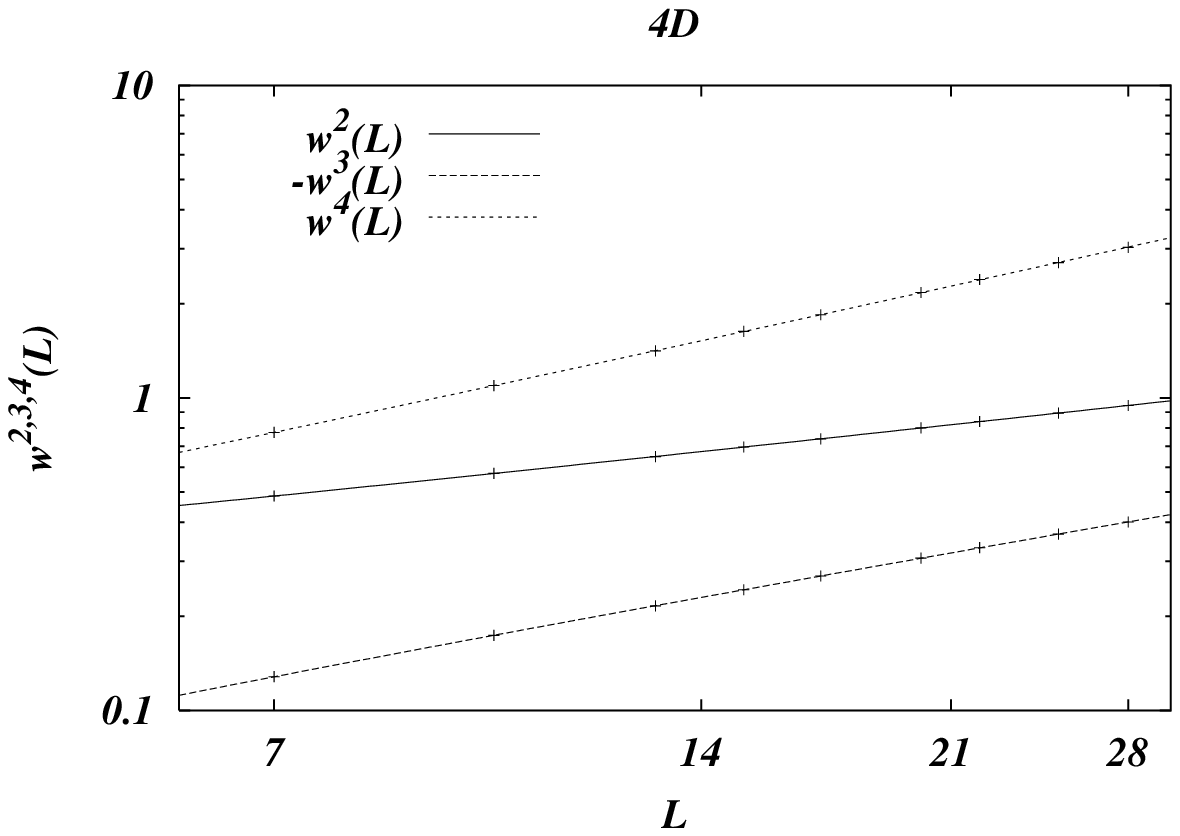}
\caption[a]{
As in figure \ref{FIG-TRE2D}, but $4D$.
\protect\label{FIG-TRE4D}}
\end{figure}

In $4D$ we use data for $L$ starting from $10$. Again the same three
cumulant analysis gives us

\begin{equation}
  \chi_{D=4} = 0.255 \pm 0.003\ ,\ \  \omega_{D=4} = 0.98\pm 0.09\ .
\end{equation}
We plot the rescaled $w_2$ versus the rescaled time in figure
\ref{FIG-SCA4D} and the numerical data for $w_2$, $w_3$ and $w_4$
versus $L$ and the best fit in \ref{FIG-TRE4D}.

Here $\chi_R=\frac27\simeq 0.286$, and, again, we are sitting a ten
standard deviations away from the rational guess.

We summarize our findings in table \ref{TAB-TWO} where we give all
the best fit values of the parameters entering equation
(\ref{E-3FIT}).

Two more observations are interesting. In first we find $\omega\simeq
1$ for all the $D$ values we investigate. In second the pre-factor of
the scaling corrections increases with $D$: we find $B_2(D=2)\simeq
0.08$, $B_2(D=3)\simeq 0.25$, $B_2(D=4)\simeq 0.37$.

\begin{table}
\centering
\begin{tabular}{|c|c|c|c|c|c|c|c|c|}
\hline
   & $A_2$ & $B_2$ & $A_3$ & $B_3$ & $A_4$ & $B_4$ & $\omega$ & $\chi$\\
\hline
2D&0.116(1)&0.08(1)&0.0105(1)&-0.8(1)&0.042(1)&-0.23(3)&1.1(3)&0.393(3)\\
\hline
3D&0.149(1)&0.25(2)&0.0226(1)&-0.9(1)&0.074(1)&0.28(1)&0.98(8)&0.3135(15)\\
\hline
4D&0.170(1)&0.37(3)&0.0321(2)&-0.7(1)&0.100(1)&0.46(4)&0.98(9)&0.255(3)\\
\hline
\end{tabular}
\caption[a]{
Best fit values of the parameters entering equation (\ref{E-3FIT}).
\protect\label{TAB-TWO}}
\end{table}

Our recent are not incompatible with the recent ones of \cite{KIM},
but our small error bars allow us to reach precise conclusions.
Also the comparison with the exponents found in \cite{ALANISETAL} is
fair: we have stressed the length of our runs, in order to be able to
give a clean estimate of the time asymptotic behavior, so that our
result is hopefully unbiased. 

\section{Conclusions}

The numerical technique we have introduced works well, and has allowed
us to run very precise numerical simulations with a limited amount of
computer time (a few months of Pentium II processor).

We have been able to determine critical exponents of the KPZ
universality class with high accuracy. We have falsified the guess
that the exponents are simple rational numbers. It is also unambiguous
from our data that the upper critical dimension is larger than $4$
(as opposed to the claims of \cite{LASKIN}).

Thanks to our precise measurements (and fitting together the first $3$
non trivial moments of $h$) we have also been able to determine the
exponent $\omega$ of the first non-leading scaling corrections.  It is
interesting to notice that the estimated value of $\omega$ is always
very close to $1$, independent from the dimensionality of the system.

The next interesting step, following the approach of \cite{CAMAPI},
would be to try and implement a systematic Monte Carlo Renormalization
Group: the Multi-Surface Coding technique we have discussed could be a
very important ingredient of such a development.

\section*{Acknowledgments}
For the numerical simulations described here we have used the {\em
Kalix2} parallel computer (built on Pentium II chips), funded by
Italian MURST 1998 COFIN.  We thank Claudio Castellano and Matteo
Marsili for interesting discussions, and Marcel den Nijs and Kay Wiese
for a relevant correspondence.

\end{document}